# Unusual pressure effects on the superconductivity of indirectly electron-doped $(Ba_{1-x}La_x)Fe_2As_2$ epitaxial films

Takayoshi Katase,[1,4] Hikaru Sato,[2] Hidenori Hiramatsu,[2,3] Toshio Kamiya,[2,3] and Hideo Hosono[1,2,3,*]

[1] Frontier Research Center, Tokyo Institute of Technology, Mailbox S2-13, 4259 Nagatsuta-cho, Midori-ku, Yokohama 226-8503, Japan

[2] Materials and Structures Laboratory, Tokyo Institute of Technology, Mailbox R3-1, 4259 Nagatsuta-cho, Midori-ku, Yokohama 226-8503, Japan

[3] Materials Research Center for Element Strategy, Tokyo Institute of Technology, Mailbox S2-16, 4259 Nagatsuta-cho, Midori-ku, Yokohama 226-8503, Japan

[4] Present address: Research Institute for Electronic Science, Hokkaido University, Sapporo 001-0020, Japan

[*]Corresponding author. E-mail: hosono@msl.titech.ac.jp

## Abstract

Applying an external pressure to indirectly electron-doped 122-type $(Ba_{1-x}La_x)Fe_2As_2$ epitaxial films enhances the superconducting critical temperature ($T_c$) up to 30.3 K. Different from the other family compounds, the $T_c$ is enhanced not only in the under-doped region but also in the optimally doped and over-doped regions. Narrowing of the superconducting transition width and an increase in the carrier density take place simultaneously in the whole doping region, except at the heavily over-doped limit. This characteristic is unique to and observed only in $(Ba_{1-x}La_x)Fe_2As_2$, in which the La doping is stabilized via non-equilibrium growth of the vapor phase epitaxy, among the 122-type iron-based superconductors, $AFe_2As_2$ ($A$ = Ba, Sr, and Ca).

Since the first report on an iron-based superconductor,[1] it has been recognized that applying external pressure is one of the predominant ways to enhance the superconducting critical temperature ($T_c$).[2,3] An iron pnictide $BaFe_2As_2$ has a layered crystal structure composed of alternating stacks of alkaline earth ($A$ = Ba, Sr, and Ca) and FeAs layers, called a '122-type structure', as shown in Fig. 1. Undoped $BaFe_2As_2$ is an antiferromagnetic metal and does not exhibit a superconducting transition under an ambient pressure.[4] However, it exhibits superconductivity with a $T_c$ up to 34 K by applying external pressures of 1–3 GPa,[5,6] along with suppression of the antiferromagnetism. This crystal structure has two major doping sites that can induce superconductivity under an ambient pressure. Substitution with a different transition metal at the Fe site is called 'direct-doping' and substitution with a different ion at the $A$ or As site is called 'indirect-doping'. This is because the Fermi level is mainly formed of Fe $3d$ orbitals. So far, pressure effects on the superconductivity of $BaFe_2As_2$ single crystals have been examined for Co-doping (direct electron-doping at the Fe sites),[7] K-doping (indirect hole-doping at the Ba sites),[8–10] and P-doping (indirect isovalent-doping at the As sites)[11] for wide ranges of doping concentrations. For the series of Co-, K- and P-doped $BaFe_2As_2$ crystals, the $T_c$ was only enhanced in the under-doped regions and never in the optimally doped and over-doped regions. This result is consistent with experimental results, showing that an external pressure significantly affected the superconductivity in the under-doped regions. It had very little effect in the over-doped region where the magnetic instability was completely suppressed.[12]

Indirect doping in 122-type $AFe_2As_2$ was reported with only hole-doping (K at the $A$ sites) and isovalent-doping (P at the As sites), but not with electron-doping, and the $T_c$ was limited to 38 K.[13] More recently, high $T_c$ values up to 49 K[14] were demonstrated in indirectly electron-doped (rare earth (*R*) metals at the *A* sites) 122-type CaFe$_2$As$_2$ single crystals grown using a flux method.[15] The reported $T_c$ is the highest among the 122-type *A*Fe$_2$As$_2$ series. On the other hand, we recently succeeded in indirect *R*- (electron-) doping (*R* = La – Nd) at the Ba sites in 122-type BaFe$_2$As$_2$ using a film-growth process. This was thermodynamically unstable under an ambient pressure and demonstrated bulk superconductivity with a maximum $T_c$ = 22 K with La doping.[16,17] This unstable phase was stabilized using the highly non-equilibrium nature of the thin film growth via a vapor phase. The maximum $T_c$ of the (Ba$_{1-x}$R$_x$)Fe$_2$As$_2$ films was much lower than that of the *R*-doped CaFe$_2$As$_2$ single crystals. According to the scenario of the above *R*-doped CaFe$_2$As$_2$ crystals (smaller *R* doping such as Pr led to a higher $T_c$), we expected that the external pressure would induce lattice shrinkage, similar to the *R*-doped CaFe$_2$As$_2$, and then enhance the $T_c$ of the (Ba$_{1-x}$La$_x$)Fe$_2$As$_2$ epitaxial films.

In this Rapid Communication we have examined the pressure effects up to 3.5 GPa on the superconductivity of indirectly electron-doped 122-type (Ba$_{1-x}$La$_x$)Fe$_2$As$_2$ epitaxial films with a wide range of *x* values from the under-doped (*x* = 0.08) to the heavily over-doped regions (*x* = 0.30). This is different from the previously reported results on other doped modes of the 122-type *A*Fe$_2$As$_2$. The (Ba$_{1-x}$La$_x$)Fe$_2$As$_2$ films exhibited a large enhancement of the $T_c$, as well as narrowing of the transition width, in the wide doping region from the under- to the optimal- and over-doped regions. This distinct characteristic is discussed in relation to the changes in the electron transport properties.

Here, 200-nm-thick epitaxial films of (Ba$_{1-x}$La$_x$)Fe$_2$As$_2$ with *x* = 0.08, 0.09, 0.13, 0.18, 0.21, and 0.30 were prepared on MgO (001) single-crystal substrates by pulsed laser deposition. Details on the procedures and sample quality are reported in Ref. [16]. The superconducting properties at 0 GPa are summarized in Fig. S1 in the Supplemental Material.[18] Each film was patterned in a six-terminal Hall-bar structure using photolithography and Ar ion milling, followed by the formation of Au electrode pads using sputtering and a lift-off process.

High-pressure experiments were performed using a piston-cylinder-type high-pressure cell. The Hall-bar samples were introduced into a Teflon cell together with a liquid-pressure-transmitting medium (Daphne oil 7474), which solidifies at 3.7 GPa at room temperature.[19] Thus, hydrostatic external pressures of 0–3.5 GPa were applied in this study. The pressure in the cell was calibrated using the measured $T_c$ of 99.99% pure Sn wires, located in the vicinity of the sample. A Cernox thermometer, attached to the body of the cell, was used to determine the sample temperature ($T$). The electrical resistivity $\rho$ ($\rho_{xx}$) and Hall resistivity $\rho_{xy}$ were measured using the four-probe technique. The electrical current flowed along the *a*-axis in the film plane. Magnetic fields up to 9 T were applied parallel to the *c*-axis, normal to the film plane. After the pressure was released we confirmed that the $T_c$, $\rho_{xx}$, and $\rho_{xy}$ recovered the initial ambient state, guaranteeing no pressure-induced degradation of the films.

Figure 2(a) shows the $\rho$–$T$ curves of the optimally doped $(Ba_{0.87}La_{0.13})Fe_2As_2$ epitaxial film measured under external pressures up to $P$ = 3.2 GPa. The $\rho$ at the normal-state monotonically decreased and the $T_c$ shifted toward higher $T$ as $P$ increased. The $\rho$–$T$ curves in the wide $T$ range up to 200 K are explained well with a simple linear law of $\rho_{fit}$ = $AT^n + \rho_0$ with the exponent $n$ = 1, where $\rho_0$ is the residual resistivity (see the solid straight lines for the fitting results in Fig. 2(a)). For $P$ = 0 GPa, the $\rho$–$T$ curve followed the linear relation well into the high-$T$ region, but deviated, showing an upturn at $T$ lower than $T_{min}$ = 120 K (the solid triangle of the $\rho$–$T$ curve at 0 GPa in Fig. 2(a)). With an increase in $P$ the deviations from the linear relations became smaller and the $T_{min}$ shifted to lower $T$. The upturn in $\rho$ is attributed to electron scattering by the La dopants as recently discussed in Ref. [17]. Compared with the results of indirectly Co-doped films, which do not exhibit any resistivity upturn, the amount of the Fe impurity phase and the crystalline qualities are almost the same in both the La- and Co-doped films.[16,20] Therefore, the electron scattering cannot be explained by external effects such as Fe impurity and different crystalline qualities. In addition, the La-doped film did not show any magnetoresistance, which implies that the electron scattering with the La dopants cannot be explained by magnetic scattering and should be attributed to disordering of the charges at the (Ba,La) sites.[17] This result indicates that the electron scattering was significantly suppressed by applying $P$. This will be further discussed below. The $P$ dependences of the onset of $T_c$ ($T_c^{onset}(P)$) and the superconducting transition width ($\Delta T_c(P)$) (defined by $\Delta T_c = T_c^{onset} - T_c^{zero}$) are summarized in Figs. 2(b) and (c), respectively. The $T_c^{onset}$ was extremely sensitive to $P$ and increased from 22.6 to 30.3 K at 2.8 GPa, leveling off at $P = 3.2$ GPa. We also confirmed that when a magnetic field was applied at $P = 3.2$ GPa, the superconducting transition became broader and $T_c$ decreased (see Fig. S2 in the Supplemental Material[18] for the magnetic field dependence), further supporting the superconductivity. With an increase in $P$, the $\Delta T_c$ also narrowed from 3.4 to 1.4 K, along with the enhancement of $T_c$. It should be noted that all the data previously reported on Co-, K- and P-doped BaFe$_2$As$_2$ crystals[7–12] indicated that $T_c$ enhancement was observed only in the under-doped region and never in the optimally doped region.

Figures 3 (A) show the $\rho$–$T$ curves for the under-doped ($x = 0.08$ and $0.09$), the over-doped ($x = 0.18$ and $0.21$), and the heavily over-doped ($x = 0.30$) epitaxial films

under various pressures (see Fig. S3 in the Supplemental Material[18] for the $\rho$–$T$ curves in the wider $T$ range). Upon applying $P$, the normal-state $\rho$ for both the under-doped and the over-doped films decreased and all of their $T_c^{onset}$ shifted to higher $T$. On the other hand, although the heavily over-doped film also exhibited a decrease in $\rho$, it did not transit to a superconducting phase for $P$ up to 3.5 GPa. Figure 3(f) summarizes the pressure effects on $T_c^{onset}(P)$ for the under-doped and the over-doped films and the inset represents the change in $\Delta T_c(P)$. For the under-doped film with $x = 0.08$, a clear saturation of $T_c^{onset}$ was not observed up to 3.4 GPa, while the films with a higher $x$ exhibited saturation of the $T_c^{onset}$ at $P \sim 3$ GPa. Since the film with $x = 0.08$ showed a resistivity anomaly at $P = 0$ GPa (see the vertical line in Fig. S3(a) in the Supplemental Material[18] for the $P$ dependence of the $\rho$–$T$ curves of the $x = 0.08$ film), caused by the coexistence of the magnetic order, the suppression of the resistivity anomaly with pressure contributed to the enhancement of $T_c$ in the higher $P$ region over 3 GPa via suppression of the magnetic order and reduction of electron scattering.

Figure 4(a) outlines the electronic phase diagram plotting the maximum $T_c^{onset}$ ($T_c^{max}$) under high pressures in comparison with the $T_c^{onset}$ at $P = 0$ GPa. The superconducting dome became larger with a higher $T_c^{onset}$ upon application of a higher $P$. It should be noted that an increase in $T_c^{onset}$ with pressure was observed for all films. The highest enhancement factor (Fig. 4(b)), defined by $T_c(P)/T_c(0)$, was obtained for the under-doped film with $x = 0.08$ (2.7 at $P = 3.4$ GPa) and the lowest enhancement factor for the optimally doped film with $x = 0.13$ was 1.4 at 3.2 GPa. The larger enhancement for the under-doped film originated from the existence of a magnetic instability at $P = 0$ GPa. The $T_c(P)/T_c(0)$ value became smaller as $x$ increased and was a minimum for the optimal value of $x = 0.13$. Almost the same $T_c(P)/T_c(0)$ values were obtained between the under-doped film with $x = 0.09$ and the over-doped film with $x = 0.21$. For the other 122-type BaFe$_2$As$_2$ crystals, such as those that were Co-, K- and P-doped,[7–11] the $T_c(P)/T_c(0)$ were nearly one or less than one (Fig. 4(b)). Compared with the pressure phase diagrams of the other 122-type BaFe$_2$As$_2$ crystals (Fig. 4(c)), the widening of the phase diagram toward the higher $T_c$ region under pressure over the whole doping region was a unique feature, observed only in the indirectly electron-doped (Ba$_{1-x}$La$_x$)Fe$_2$As$_2$ epitaxial films among the 122-type $A$Fe$_2$As$_2$. Here we should point out that, although the epitaxial films are affected by uniaxial pressures originating from the lattice mismatching between the substrates and the films (epitaxial strain), this is not essential for the present results. To discuss the epitaxial strain, we examined the pressure effects on the $T_c^{onset}$ for the optimally doped Ba(Fe$_{1-x}$Co$_x$)$_2$As$_2$[20] and BaFe$_2$(As$_{1-x}$P$_x$)$_2$[21] epitaxial films grown on MgO (001) single crystals and compared them with that of the (Ba$_{1-x}$La$_x$)Fe$_2$As$_2$ films (see Fig. S4 in the Supplemental Material[18]). The former two films had in-plane lattice mismatches ($-6.2 \pm 0.3\%$) similar to those ($-5.9\%$) of the (Ba$_{1-x}$La$_x$)Fe$_2$As$_2$/MgO films. Irrespective of the large epitaxial strains, the $T_c$ decreased and was not enhanced for both of the samples. Similarly, optimally doped (Ba$_{1-x}$K$_x$)Fe$_2$As$_2$ films exhibited a pressure effect on the $T_c^{onset}$ but it was quite small.[22] These results suggest that the $T_c$ enhancement in the (Ba$_{1-x}$La$_x$)Fe$_2$As$_2$ epitaxial films did not originate from the epitaxial strain but from the external hydrostatic pressure.

Figures 5(a–c) show the $T$ dependences of the Hall coefficient ($R_H$) for the under-doped ($x = 0.08$), the optimally doped ($x = 0.13$) and the over-doped ($x = 0.21$) epitaxial films at different pressures. The $R_H$ values were all negative and their absolute values ($|R_H|$) increased with decreasing $T$ for all of the films. With an increasing $P$ the $R_H$ values at high $T$ were almost unchanged, while the $|R_H|$ values decreased at low $T$,

giving a smaller $T$ dependence under a higher $P$. This result implies that the $P$ induced more mobile carriers at lower $T$. For a quantitative discussion of the mobile carrier density, we should assess whether we can apply the Hall effect relation: $N_e = 1/(e \times |R_H|)$ (where $N_e$ is the mobile carrier density and $e$ is the elementary electric charge) to BaFe$_2$As$_2$ because BaFe$_2$As$_2$ has a complex Fermi surface composed of three hole bands and two electron bands. In the case of directly electron-doped (Co-doped) BaFe$_2$As$_2$, it has been reported that the hole contribution to the carrier transport was negligible and the above relation provides reasonable $N_e$ at low $T$ in most of the doping regions.[23] In Fig. S5 in the Supplemental Material,[18] we measured $R_H$ at $P = 0$ GPa and estimated the values of $N_e$ per Fe site ($N_e$/Fe) using the equation $N_e$/Fe $= 1/(e \times |R_H|)/V/4$, where $V$ is the unit cell volume containing four Fe atoms for (Ba$_{1-x}$La$_x$)Fe$_2$As$_2$ epitaxial films. Fig. S5(b) in the Supplemental Material[18] shows that the normalized $N_e$/Fe (plotting against $x/2$ for (Ba$_{1-x}$La$_x$)Fe$_2$As$_2$ and $x$ for Ba(Fe$_{1-x}$Co$_x$)$_2$As$_2$[24]) follows the same relationship between the directly Co-doped and the indirectly La-doped BaFe$_2$As$_2$. This result supports the idea that the same conclusion can be applied also for (Ba$_{1-x}$La$_x$)Fe$_2$As$_2$. Therefore, we directly estimated $N_e$ as $1/(e \times |R_H|)$ (Fig. 5(d)). $N_e$ at 35 K gradually increased with $P$ for all of the films. The relationships between the estimated $N_e$ and $T_c^{onset}$ for the three kinds of films from the under- to the over-doped regions are superimposed in the inset of Fig. 5(d). The $x = 0.21$ film exhibited a $T_c^{onset} - N_e$ curve that was very different from those of the other films. A universal relation such as a dome-like single $T_c^{onset} - N_e$ relation was not observed. This trend implies that the $T_c^{onset}$ of the (Ba$_{1-x}$La$_x$)Fe$_2$As$_2$ epitaxial films under a high pressure was not determined just by $N_e$. The estimated Hall mobilities ($\mu$) were independent of pressure, shown in the insets of Fig. 5(a–c). These results indicate that $P$ increases $N_e$ but does not affect the mobility and does not disturb the carrier transport. This consideration is consistent with the indirect La doping at the Ba sites in $BaFe_2As_2$. We considered that the reduction of electron scattering and the increase in $N_e$ induced by $P$ would be a plausible origin for the $T_c$ enhancement because a large enhancement of $T_c^{onset}$ shows a good correlation with the suppression of the $\rho$ upturn (shown in Figs. 2(a) and S3) and the increase in $N_e$ at low $T$ (shown in Fig. 5).

In summary, we observed a unique pressure phase diagram in indirectly electron-doped 122-type $(Ba_{1-x}La_x)Fe_2As_2$ epitaxial films. The enhancement of $T_c$ in all of the doping regions along with narrowing of $\Delta T_c$ was associated with the reduction of electron scattering and the increase in the carrier density caused by lattice shrinkage, which optimizes its crystal and electronic structure to achieve higher $T_c$ and sharper $\Delta T_c$ in $(Ba_{1-x}La_x)Fe_2As_2$ films.

**Acknowledgments**

This work was supported by the Japan Society for the Promotion of Science (JSPS), Japan, through the "Funding Program for World-Leading Innovative R&D on Science and Technology (FIRST Program)" and MEXT Element Strategy Initiative to Form Core Research Center.

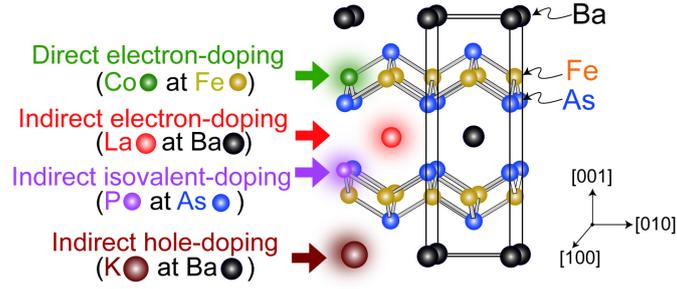

FIG. 1. (Color online) Crystal structure and doping modes of the layered 122-type BaFe$_2$As$_2$. Solid lines show a unit cell. The doping sites that induce superconductivity are categorized into two modes. One is 'indirect-doping' for doping at sites other than the Fe sites, and the other is 'direct-doping' for doping at the Fe sites.

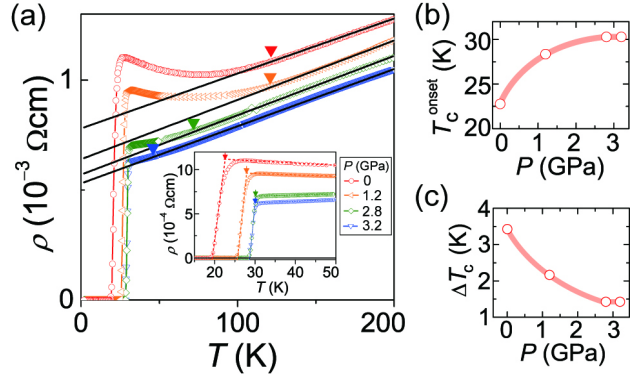

FIG. 2. (Color online) Pressure effects on the superconductivity of the optimally doped (Ba$_{0.87}$La$_{0.13}$)Fe$_2$As$_2$ epitaxial films. (a) $\rho$–$T$ curves were measured under various pressures from an ambient pressure up to $P$ = 3.2 GPa. The solid lines are the fitting results to the power law, $\rho_{\text{fit}}$ = A$T^n$ +$\rho_0$, with $n$ = 1. The solid triangles indicate the positions of $T_{\text{min}}$. An expanded view of the $\rho$–$T$ curves at low $T \leq$ 50 K is shown in the inset. The arrows indicate the positions of $T_c^{\text{onset}}$. Summary of the variation of (b) $T_c^{\text{onset}}$ and (c) $\Delta T_c$ against pressure.

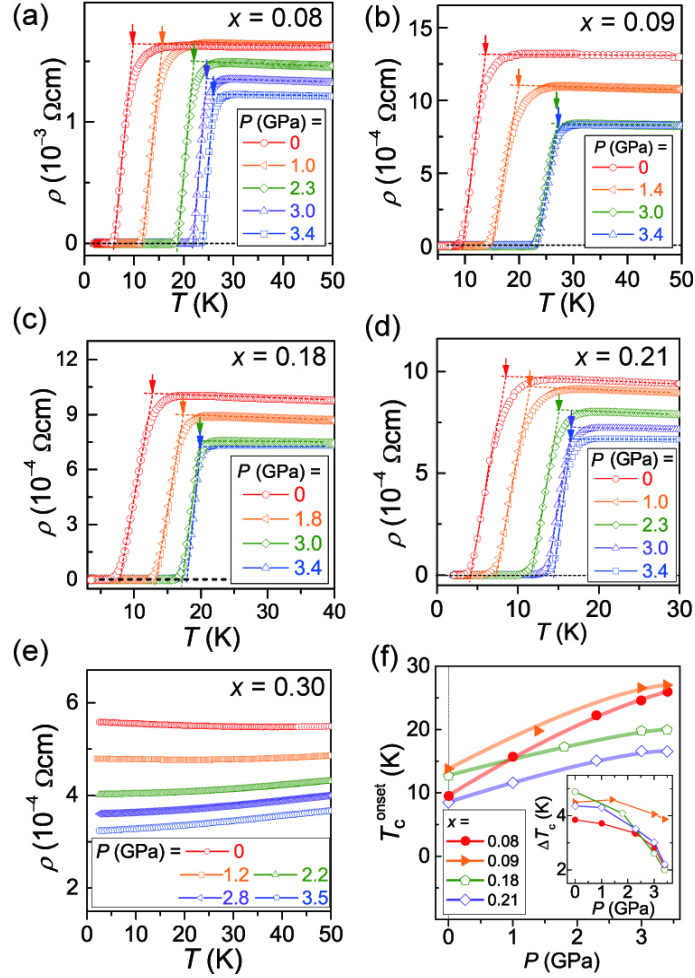

FIG. 3. (Color online) (A) $\rho$–$T$ curves measured under various pressures for $(Ba_{1-x}La_x)Fe_2As_2$ epitaxial films that are under-doped [(a) $x = 0.08$ and (b) $0.09$], over-doped [(c) $x = 0.18$ and (d) $0.21$], and heavily over-doped [(e) $x = 0.30$]. The arrows indicate the positions of $T_c^{onset}$. The applied pressures are indicated in each panel. (f) Pressure dependences of $T_c^{onset}$ ($T_c(P)$) for the under-doped and over-doped films. The inset shows the changes in $\Delta T_c$ ($\Delta T_c(P)$) as a function of $P$. Open and solid symbols represent $T_c(P)$ and $\Delta T_c(P)$ for under-doped and over-doped films, respectively.

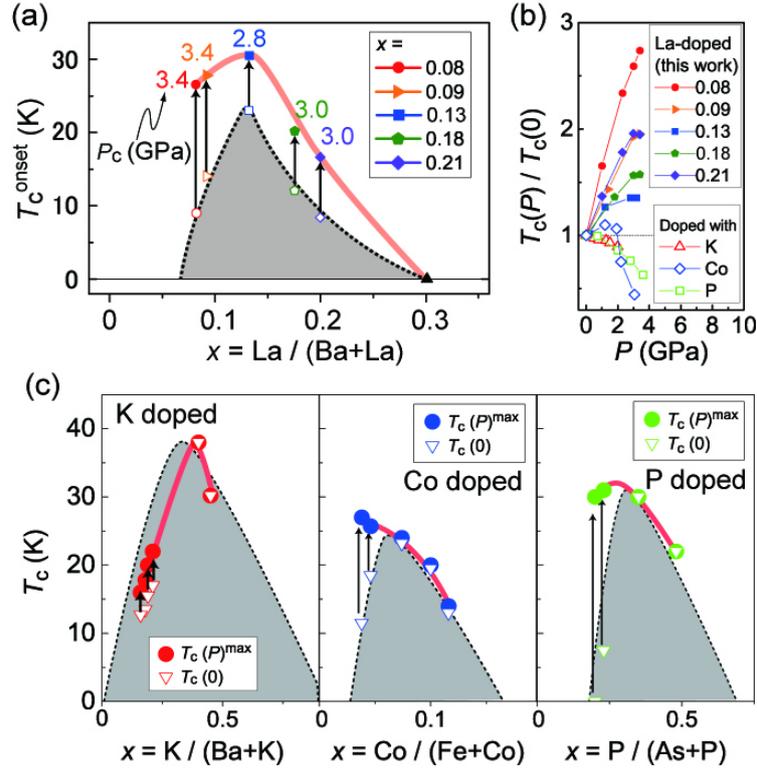

FIG. 4. (Color online) (a) High-pressure electronic phase diagram of $(Ba_{1-x}La_x)Fe_2As_2$ epitaxial films in comparison with that under an ambient pressure. Solid and open symbols show the maximum $T_c^{onset}$ ($T_c^{max}$) obtained under high pressures and those under an ambient pressure, respectively. The critical pressures ($P_c$), where $T_c^{max}$ was obtained are indicated in the figure. (b) Enhancement factors defined by $T_c(P)/T_c(0)$ as a function of $P$ for $(Ba_{1-x}La_x)Fe_2As_2$ epitaxial films. Those of the optimally, K (indirect hole)-, P (indirect isovalent)-, and Co (direct electron)-doped $BaFe_2As_2$ single crystals are shown for comparison.[7,8,11] (c) Pressure phase diagrams of $(Ba_{1-x}K_x)Fe_2As_2$ (indirect hole-doped), $BaFe_2(As_{1-x}P_x)_2$ (indirect isovalent-doped), and $Ba(Fe_{1-x}Co_x)_2As_2$ (direct electron-doped) single crystals.[7-11] Typical electronic phase diagrams under an ambient pressure are shown by the dark areas. Solid and open symbols indicate $T_c^{max}$ under external pressures ($T_c(P)^{max}$) and $T_c$ at 0 GPa, respectively.

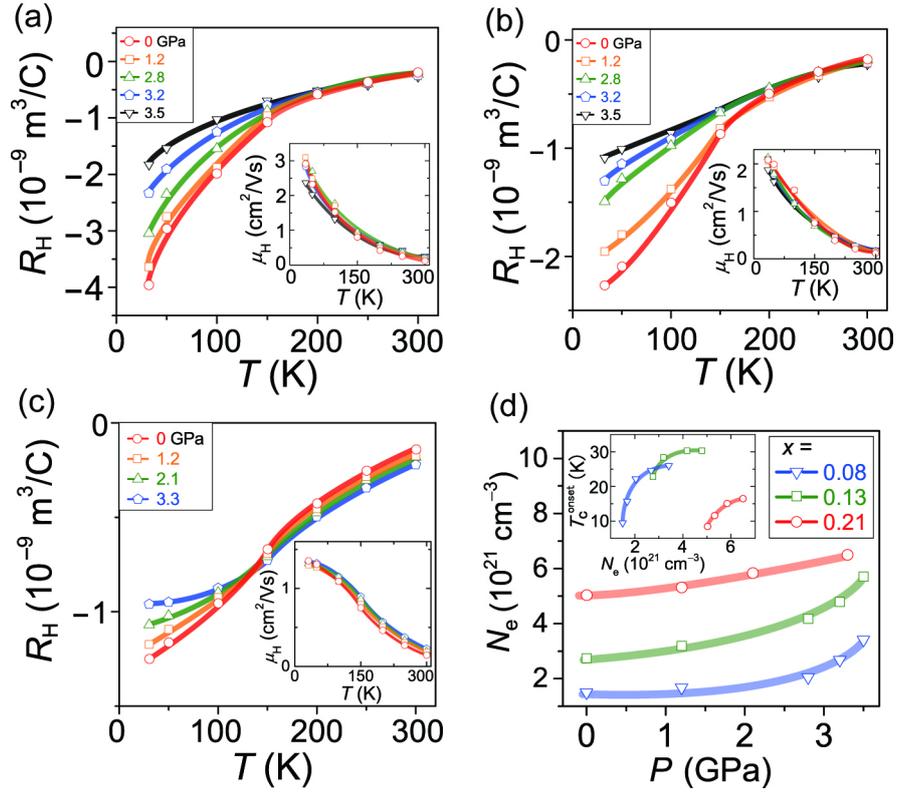

FIG. 5. (Color online) (a–c) Temperature dependence of the Hall coefficients ($R_H$) for $(Ba_{1-x}La_x)Fe_2As_2$ epitaxial films with $x = 0.08$ ((a) under-doped), $x = 0.13$ ((b) optimally doped), and $x = 0.21$ ((c) over-doped) under high pressures. The applied pressures are indicated on the upper left of each panel. The inset shows the temperature dependence of the Hall mobilities ($\mu_H$) under various pressures. (d) Pressure dependences of the carrier density ($N_e$) at 35 K. The relationship between $T_c^{\text{onset}}$ and $N_e$ is summarized in the inset.

Supplemental Material for

**"Unusual pressure effects on the superconductivity of indirectly electron-doped (Ba$_{1-x}$La$_x$)Fe$_2$As$_2$ epitaxial films"**

Takayoshi Katase,[1,4] Hikaru Sato,[2] Hidenori Hiramatsu,[2,3] Toshio Kamiya,[2,3] and Hideo Hosono[1,2,3,*]

[1] Frontier Research Center, Tokyo Institute of Technology, Mailbox S2-13, 4259 Nagatsuta-cho, Midori-ku, Yokohama 226-8503, Japan

[2] Materials and Structures Laboratory, Tokyo Institute of Technology, Mailbox R3-1, 4259 Nagatsuta-cho, Midori-ku, Yokohama 226-8503, Japan

[3] Materials Research Center for Element Strategy, Tokyo Institute of Technology, Mailbox S2-16, 4259 Nagatsuta-cho, Midori-ku, Yokohama 226-8503, Japan

[4] Present address: Research Institute for Electronic Science, Hokkaido University, Sapporo 001-0020, Japan

[(*)] E-mail: hosono@msl.titech.ac.jp

## I. Superconductivity of (Ba$_{1-x}$La$_x$)Fe$_2$As$_2$ epitaxial films under an ambient pressure

**Figure S1** shows an electronic phase diagram of (Ba$_{1-x}$La$_x$)Fe$_2$As$_2$ epitaxial films under an ambient pressure. The closed circles refer to the onset of $T_c$ ($T_c^{onset}$) and the doping concentration ($x$) relation of the films used in the present study. The open circles represent the resistivity anomaly temperature ($T_{anom}$), which is observed only in the under-doped films with $x = 0.08$. The $\rho$–$T$ curves of the (Ba$_{1-x}$La$_x$)Fe$_2$As$_2$ epitaxial films are shown in the inset.

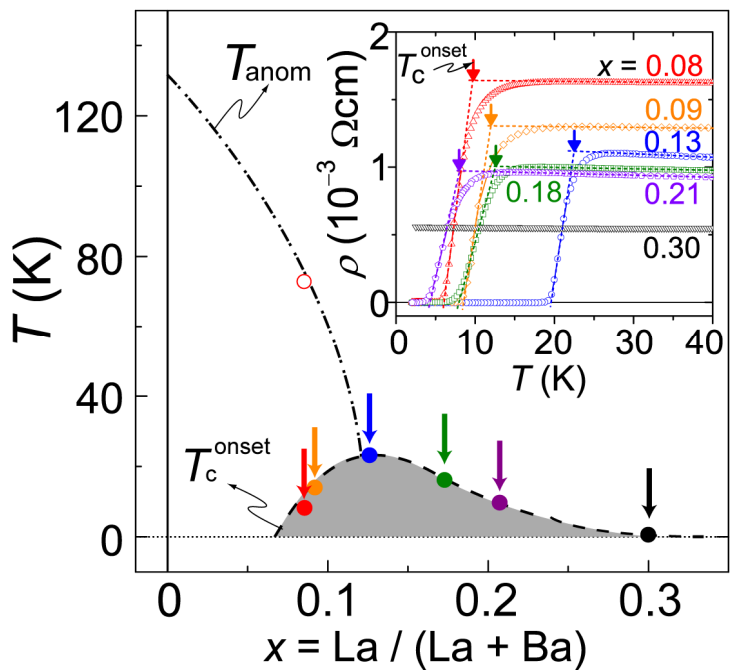

**Figure S1.** Electronic phase diagram of (Ba$_{1-x}$La$_x$)Fe$_2$As$_2$ epitaxial films under an ambient pressure. The arrows indicate the positions of the $T_c^{onset}$ for the (Ba$_{1-x}$La$_x$)Fe$_2$As$_2$ epitaxial films with $x = 0.08$, 0.09, 0.13, 0.18, 0.21, and 0.30. The inset shows the $\rho$–$T$ curves of these films.

## II. Magnetic field dependence for the $\rho$–$T$ curves for the optimally doped $(Ba_{0.87}La_{0.13})Fe_2As_2$ epitaxial films at $P = 0$ and 3.2 GPa

Magnetic fields up to 9 T were applied parallel to the $c$-axis. It was observed that both the $T_c^{onset}$ and $T_c^{zero}$ decreased with increasing magnetic field ($H$).

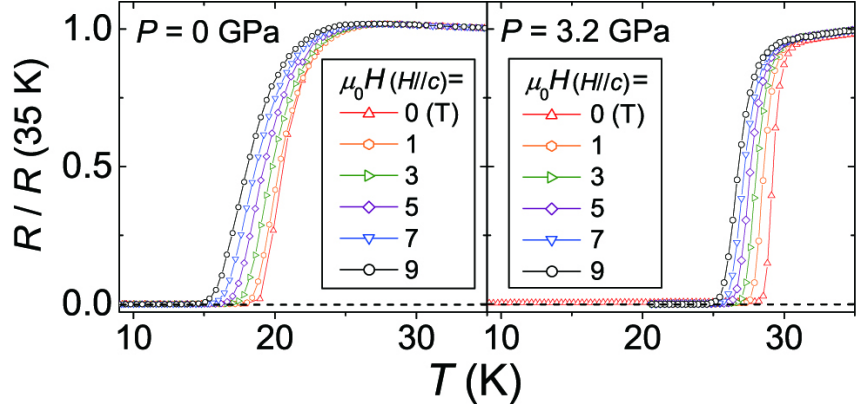

**Figure S2.** Magnetic field ($H$) dependence of the $\rho$–$T$ curves for the optimally doped $(Ba_{0.87}La_{0.13})Fe_2As_2$ epitaxial film under an ambient pressure (left panel) and a high pressure of 3.2 GPa (right panel).

## III. Pressure dependence of the $\rho$–$T$ curves for the under-doped and over-doped $(Ba_{1-x}La_x)Fe_2As_2$ epitaxial films

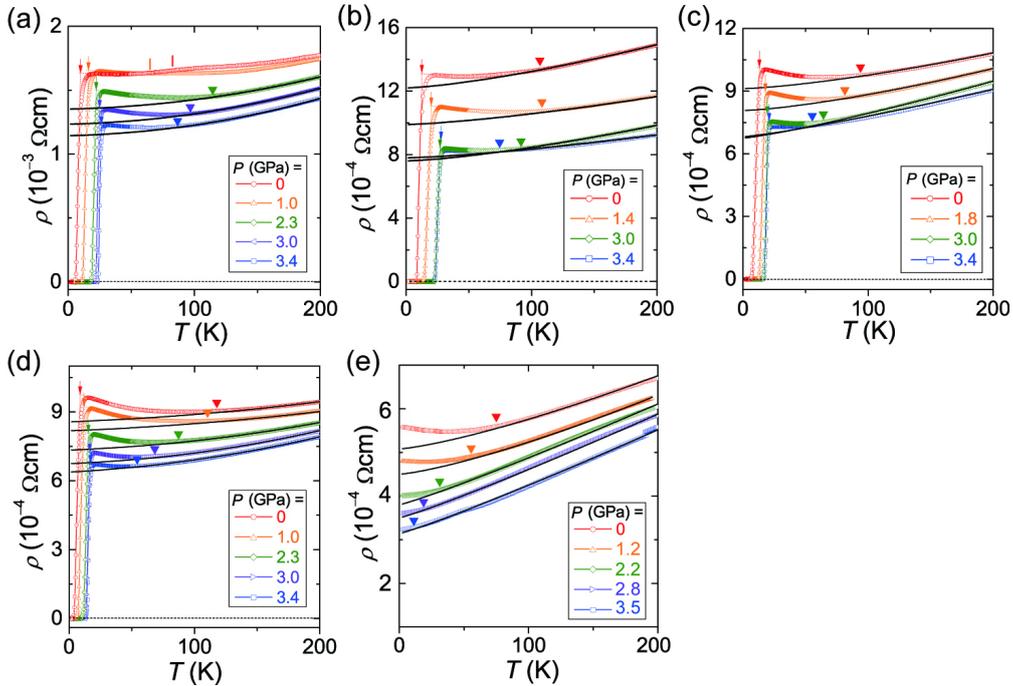

**Figure S3.** $\rho$–$T$ curves measured under various pressures for $(Ba_{1-x}La_x)Fe_2As_2$ epitaxial films that are under-doped [(a) $x = 0.08$ and (b) 0.09], over-doped [(c) $x = 0.18$ and (d) 0.21], and heavily over-doped [(e) $x = 0.30$]. The arrows and the triangles indicate the positions of $T_c^{onset}$ and $T_{min}$, respectively. The vertical lines in (a) indicate the positions of the resistivity anomaly temperature. The applied pressures are indicated in each panel. The expanded views near $T_c$ are shown in Fig. 3(a – e) in the main text.

## IV. Pressure effects on the superconductivity of directly cobalt-doped and indirectly phosphorus-doped BaFe₂As₂ epitaxial films on MgO (001) substrates

Pressure transmittance media for the BaFe$_2$(As$_{1-x}$P$_x$)$_2$ epitaxial films were changed to a Fluorinert mixture of FC70:FC77 with 1:1, because soaking a BaFe$_2$(As$_{1-x}$P$_x$)$_2$ film in Daphne 7474 and 7373 severely degraded the $T_c$, even under an ambient pressure. Fig. S4 shows the $\rho$–$T$ curves under varied pressures. The insets summarize the pressure evolution of $T_c^{onset}$. The pressure decreased $T_c^{onset}$ and $T_c^{zero}$ and broadened $\Delta T_c$.

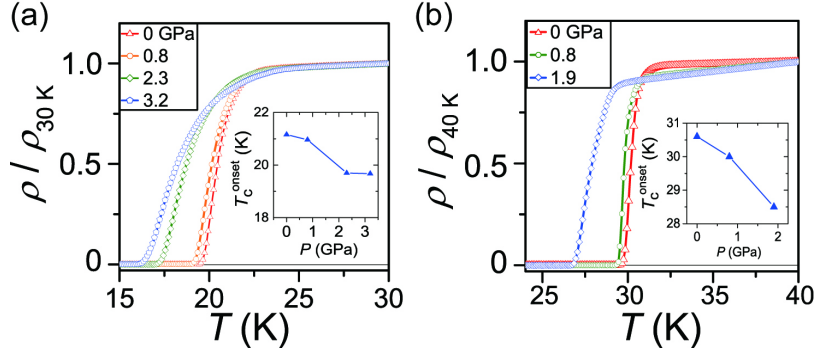

**Figure S4.** Pressure effects of the $\rho$–$T$ curves for (a) Ba(Fe$_{1-x}$Co$_x$)$_2$As$_2$ and (b) BaFe$_2$(As$_{1-x}$P$_x$)$_2$ epitaxial films on MgO (001) single-crystal substrates. The applied pressures are indicated on upper left in each figure. The insets show the pressure dependences of $T_c^{onset}$.

## V. Estimation of the carrier densities for (Ba$_{1-x}$La$_x$)Fe$_2$As$_2$ epitaxial films using Hall effect measurements

As shown in Figure S5 (a), the Hall coefficient ($R_H$) decreased with a decreasing temperature. The absolute values of $R_H$ decreased with an increase in the doping concentration ($x$) at the same temperature. Figure S5 (b) represents the $x$ dependence of the carrier densities ($N_e$), estimated from $R_H$ and extrapolated to $T = 0$. The previously reported $N_e$ – $x$ data for Ba(Fe$_{1-x}$Co$_x$)$_2$As$_2$ single crystals are superimposed in Figure S5(b).[S1] Here, the doping concentrations are normalized with the doped carriers per Fe atom ($x/2$ for La doping and $x$ for Co-doping). Both the $N_e$ – $x$ plots followed the single line, indicating that the mobile electron densities are almost the same as those expected from $x$.

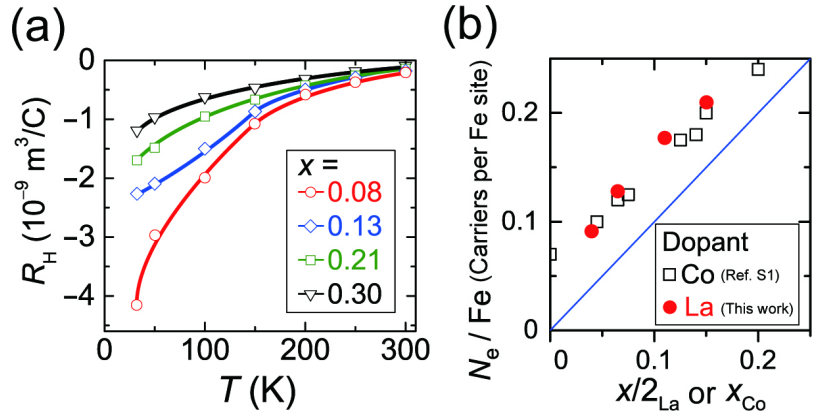

**Figure S5.** (a) Temperature dependence of the Hall coefficient $R_H$ under an ambient pressure. The circles, diamonds, squares and triangles represent the $R_H$ of (Ba$_{1-x}$La$_x$)Fe$_2$As$_2$ epitaxial films with $x$ = 0.08, 0.13, 0.21, and 0.30, respectively. (b) Doping concentration dependences of carrier density ($N_e$).